# Significant enhancement of upper critical fields by doping and strain in Fe-based superconductors

C. Tarantini[1*], A. Gurevich[2], J. Jaroszynski[1], F. Balakirev[3], E. Bellingeri[4], I. Pallecchi[4], C. Ferdeghini[4], B. Shen[5], H.H. Wen[5] and D. C. Larbalestier[1]

[1]Applied Superconductivity Center, National High Magnetic Field Laboratory, Florida State University, 2031 E. Paul Dirac Dr., Tallahassee FL 32310, USA

[2]Department of Physics, Old Dominion University, Norfolk, VA 23529, USA

[3]National High Magnetic Field Laboratory, Los Alamos National Laboratory, Los Alamos, NM 87545, USA

[4]CNR-SPIN, Corso Perrone 24, 16152 Genova, Italy

[5]Institute of Physics, Chinese Academy of Sciences, Beijing, China

*e-mail: tarantini@asc.magnet.fsu.edu

We report measurements of the upper critical field $H_{c2}$(T) up to 85 Tesla on $Ba_{1-x}K_xAs_2Fe_2$ single crystals and $FeSe_{1-x}Te_x$ films tuned by doping and strain. We observed an $H_{c2}$ enhancement by more than 25 T at low temperatures for the optimally-doped $Ba_{1-x}K_xAs_2Fe_2$ as compared to the previous measurements and extraordinarily high slopes $dH_{c2}/dT$ = 250-500 T/K near $T_c$ in $FeSe_{1-x}Te_x$ indicating an almost complete suppression of the orbital pair-breaking. Theoretical analysis of $H_{c2}$(T) in the optimally doped $Ba_{1-x}K_xAs_2Fe_2$ suggests an inhomogeneous Fulde-Ferrel-Larkin-Ovchinnikov state for H//*ab* at T <10 K, and at T < 3-4K for H//*c* and <10K for H//*ab* for $FeSe_{1-x}Te_x$. The analysis also shows that $H_{c2}$ in multiband Fe based superconductor can be significantly enhanced by doping and strain by shrinking and expanding different pockets of the Fermi surface, which can be much more effective than the conventional way of increasing $H_{c2}$ by nonmagnetic impurities.

# I. INTRODUCTION

The discovery of the diverse family of Fe based superconductors[1] (FBS) in which an interplay of superconductivity and magnetism in layered semi-metals can result in transition temperatures $T_c$ up to 55K[2,3] has caused great interest. FBS also have extremely high and only moderately anisotropic upper critical fields $H_{c2}(T)$ which often significantly exceed the paramagnetic limit $B_p[T]=1.84T_c[K]$ at which the Zeeman depairing energy is greater than the binding energy of the Cooper pairs[4,5,6,7,8,9,10,11]. As a result, FBS may exhibit exotic effects at high magnetic fields, in particular, paramagnetic pair-breaking for multiband pairing[12], and the inhomogeneous Fulde-Ferrel-Larkin-Ovchinnikov (FFLO) state.[13,14,15]

The small Cooper pair sizes $\xi_0$=1-3nm of FBS indicate that the usual way of increasing $H_{c2}$ by introducing disorder in the dirty limit $\ell < \xi_0$[16] can hardly work in FBS as it would require an elastic mean free path $\ell$ close to the lattice spacing. However, the low Fermi energies and multiband electronic structure of FBS offer a new opportunity of tuning $H_{c2}$ by shifting the Fermi level through expansion or contraction of different electron and hole pockets of the Fermi surface (FS), which also facilitates the FFLO transition.[12] This feature is amplified by a doping-induced Lifshitz transition caused by the emergence of small FS pockets from bands just below the Fermi level[12] which have been revealed by *ab-initio* calculations, angular resolved photoemission (ARPES),[2,17,18,19,20] particularly for FeSe$_{1-x}$Te$_x$ which has rather small Fermi energies E$_F$ = 20-50 meV. In this case the increase of $H_{c2}$ is not just a mere consequence of the dependence of $T_c$ on doping, because the shape of $H_{c2}$(T) also changes significantly due to interplay of orbital and paramagnetic pairbreaking in different bands. In this work we provide the first experimental evidence for the significant enhancement of $H_{c2}$ in the optimally doped Ba$_{1-x}$K$_x$As$_2$Fe$_2$ at below 30 K using high-field measurements of $H_{c2}(T)$ up to 85T, which extend well beyond the previous measurements of $H_{c2}$ on Ba$_{1-x}$K$_x$As$_2$Fe$_2$[4,7,21] and FeSe$_{1-x}$Te$_x$.[8,9]

## II. MEASUREMENTS

We studied three $Ba_{1-x}K_xFe_2As_2$ crystals grown by the flux method,[22,23] one optimally doped with x=0.4 (Ba-1) and two underdoped ones with x=0.25 (Ba-2) and 0.15 (Ba-3).The samples had resistive transition midpoints at $T_c$=38.4, 28.2 and 11.7K with transition widths $\Delta T_c$=1, 1.1 and 2.6K, respectively. $FeSe_{0.5}Te_{0.5}$ films were grown by pulsed laser deposition from a target with $T_c$ =16.2K. Several c-axis and in-plane oriented films with thickness from 100 to 400nm and 2x5mm$^2$ in the ab plane were grown on single crystal $LaAlO_3$(001) substrates in high vacuum at 550°C. The films have $T_c$=17.5-18.9K, and $\Delta T_c$ from 0.8 to 1.3K. These films were characterized in Refs. 24 and 25, where the enhancement of $T_c$ by strains was related to the distortions of the (Se,Te)–Fe–(Se,Te) bond angle and length.

Magneto-transport measurements of the resistance R(H) at low fields were performed in a 9T Quantum Design physical property measurement system, while high field measurements were done within the 45T hybrid DC magnet and the 60 and 85T pulsed magnets at the National High Magnetic Field Laboratory in Tallahassee and Los Alamos. To minimize heating in pulsed fields, $Ba_{1-x}K_xFe_2As_2$ crystals were cleaved down to 2x0.3x0.1mm$^3$. A 100kHz AC current was employed and the data were analyzed using a low-noise digital lock-in technique. Since the heating effect is usually more pronounced at lower temperatures and with the field perpendicular to the largest sample surface, we verified that the up and down field sweeps in such conditions for Ba-2 do not result in noticeable hysteresis in the R(H) curves as shown in Figure 1. The almost identical transition position and the absence of any low-field tail in R(H) in the down sweeps are indicative of negligible heating effects. The sample resistance R(T,H) was measured for fields both parallel and perpendicular to the ab planes ($H_{c2}^{//}(T)$ and $H_{c2}^{\perp}(T)$, respectively) and $H_{c2}(T)$ was defined at commonly used criterion $R(T,H_{c2})=0.9R_n(T,H)$ where $R_n$ is the normal state resistance.

We chose this 90%-criterion to extrapolate $H_{c2}$ from resistive transitions in order to reduce the effect of thermal fluctuation effects and sample inhomogeneities on $H_{c2}$. It is known that different resistive criteria may affect the measured $H_{c2}$(T) temperature dependencies, as was shown for Nd-1111 pnictide single crystals.[11] Moreover, specific heat and transport measurements produce somewhat different results for $H_{c2}$.[26] This discrepancy may naturally result from the inevitable local $T_c$ inhomogeneities, which determine the transition width even in the best single chalcogenide single crystals.[26] Since the variations of $T_c$ on scales greater than the coherence length result in a distribution of local $H_{c2}$, there is no unambiguous definition for the global $H_{c2}$ in an inhomogeneous superconductor. As a result, transport and specific heat measurements probe different parts of the distribution of $H_{c2}$: magneto-transport at the onset of superconductivity probes the higher-$T_c$ part of the $H_{c2}$ distribution in regions which form a percolative cluster. The specific heat measurements probe both higher-$T_c$ and lower-$T_c$ parts of the $H_{c2}$ distribution which correspond to regions in which superconductivity could be modified by local non-stoichiometry or strains which cause deviation of the local Fe-As or Fe-(Te,Se) bond angles.[2,3] It should also be noted that fluctuation effect can change the shape of resistivity and specific heat curves in different ways.

Figure 2 shows that the in-field transitions of Ba-1 taken in the 60 T magnet with the pulse width of ~75ms are clean and sharp, indicating high quality of our samples. The transition curves R(H) obtained in the 85T magnet are only slightly noisier because of the shorter pulse width (~15ms). $H_{c2}^{\perp}(T)$ and $H_{c2}^{//}(T)$ are shown in Figure 3a for the optimally-doped and underdoped Ba$_{1-x}$K$_x$Fe$_2$As$_2$ where // and ⊥ correspond to the magnetic fields parallel and perpendicular to the *ab* planes, respectively. Here $H_{c2}^{//}(T)$ curves have downward curvatures in all the cases. By contrast, $H_{c2}^{\perp}(T)$ for the optimally doped Ba-1 and slightly underdoped Ba-2 have downward curvatures, whereas the most underdoped Ba-3 has an upward curvature in $H_{c2}^{\perp}(T)$. This behavior is consistent with the behavior of orbitally-limited $H_{c2}(T)$ for multiband s$^{\pm}$ pairing where the order parameter changes sign on different sheets of the FS[12]. As *T* decreases, the anisotropy parameter $\gamma(T)=H_{c2}^{//}(T)/H_{c2}^{\perp}(T)$ decreases from $\gamma(T_c)\approx 2$ to $\gamma(0)\approx 1$. Going from the strongly underdoped Ba-3 to the underdoped Ba-2 sample, $H_{c2}(T)$ nearly doubles, but $H_{c2}^{\perp}(T)$ flattens at low *T* and the upward curvature disappears. For the optimally doped Ba-1, $H_{c2}(0)$ is well above the 85T limit of our measurements. Normalizing $H_{c2}$ data to $T_c$ as shown in Fig. 3 reveals similar trend for Ba-1 and Ba-2 but a clearly different behavior for Ba-3 showing that $H_{c2}$ does not simply scale with $T_c$ because the shape of $H_{c2}(T)$ changes upon doping.

Figure 4 shows the in-field resistive transitions in FeSe$_{0.5}$Te$_{0.5}$ film measured in the 60T magnet. The comparatively sharp transitions at the highest magnetic field indicate good homogeneity of the sample. Shown in Figure 5 are $H_{c2}(T)$ measured on FeSe$_{0.5}$Te$_{0.5}$ films which all have $H_{c2}(T)$ with a downward curvatures for both field orientations. In particular, $H_{c2}^{//}(T)$ has very steep slopes $H_{c2}'$ near $T_c$ resulting in $H_{c2}^{//}(T_c/2)>40$ T. Shown in Figure 5c is an example of the resistance curve $R(T)$ for one of the FeSe$_{0.5}$Te$_{0.5}$ films which shifted by only 2mK at 1T, corresponding to $H_{c2}'$=500T/K (Figure 5d). Such extremely high $H_{c2}$ slopes near $T_c$ are characteristic of our strained films with $H_{c2}'$ ranging from 250 to 500T/K for $H//ab$ and from 8 to 21T/K for $H//c$ (no apparent correlation between the film thickness and $H_{c2}'$ was found). For different $H_{c2}$ criterion (e.g. $R(T,H_{c2})=0.5R_n(T,H)$) the slope still exceeds 100T/K, much larger than any value previously reported in strain-free optimally doped FeSe$_{0.5}$Te$_{0.5}$ single crystals ($H_{c2}'$=13-40T/K).[10]

### III. ANALYSIS OF THE EXPERIMENTAL DATA

We interpret our $H_{c2}(T)$ data using a model of $H_{c2}(T)$ which takes into account orbital and paramagnetic pair-breaking, and the FFLO instability in clean multiband superconductors.[12] The FFLO instability at H = $H_{c2}$ is characterized by the wave vector **Q** of spatial oscillations of the superconducting order parameter $\Psi(\mathbf{r})=\Delta(x,y)[c_1\exp(iQz) + c_2\exp(-iQz)]$ which appear spontaneously below the temperature $T_F$ along **H** (if **H** is parallel to the symmetry axis). The equation for $H_{c2}^{\perp}$ is given by

$$a_1 G_1 + a_2 G_2 + G_1 G_2 = 0 \quad (1)$$

$$G_1 = \ln t + 2e^{q^2} \operatorname{Re} \sum_{n=0}^{\infty} \int_q^{\infty} e^{-u^2} \left[ \frac{u}{n+1/2} - \frac{t}{\sqrt{b}} \tan^{-1}\left( \frac{u\sqrt{b}}{t(n+1/2)+i\alpha b} \right) \right] du \quad (2)$$

where $G_2$ is obtained substituting $\sqrt{b} \rightarrow \sqrt{\eta b}$ and $q \rightarrow q\sqrt{s}$ in $G_1$. Here $Q(T,H)$ is determined by the condition that $H_{c2}(T,Q)$ is maximum, $a_1=(\lambda_0 + \lambda_-)/2w$, $a_2=(\lambda_0 - \lambda_-)/2w$, $\lambda_-=\lambda_{11} - \lambda_{22}$, $\lambda_0=(\lambda_-^2 + 4\lambda_{12}\lambda_{21})^{1/2}$, $w=\lambda_{11}\lambda_{22} - \lambda_{12}\lambda_{21}$, $t=T/T_c$, and

$$b = \frac{\hbar^2 v_1^2 H}{8\pi \phi_0 k_B^2 T_c^2}, \quad q^2 = \frac{Q_z^2 \varepsilon_1 \phi_0}{2\pi H}, \quad \alpha = \frac{4\mu \phi_0 T_c}{\hbar^2 v_1^2}, \quad (3)$$

where $\eta=(v_2/v_1)^2$, $s=\varepsilon_2/\varepsilon_1$, $v_i$ the in-plane Fermi velocity in band $i=1,2$, $\varepsilon_i=m_i^{ab}/m_i^c$ the mass anisotropy ratio, $\phi_0$ the flux quantum, $\mu$ the magnetic moment of a quasiparticle, $\lambda_{11}$ and $\lambda_{22}$ the intraband pairing constants, $\lambda_{12}$ and $\lambda_{21}$ the interband pairing constants, and $\alpha \approx \alpha_M/1.8$ where the Maki parameter $\alpha_M = 2^{1/2} H_{c2}^{orb}/H_p$ quantifies the strength of the Zeeman pairbreaking. The fits for $H//ab$ were done assuming that the bands have the same anisotropy parameter $\varepsilon_1=\varepsilon_2=\varepsilon$ for which $H_{c2}^{//}$ is defined by Eqs. (1)-(3) with the rescaled $q \rightarrow q\varepsilon^{3/4}$, $\alpha \rightarrow \alpha\varepsilon^{1/2}$ and $b^{1/2} \rightarrow b^{1/2}\varepsilon^{1/4}$ in $G_1$ and $(\eta b)^{1/2} \rightarrow (\eta b)^{1/2} \varepsilon^{1/4}$ in $G_2$.[12]

Shown in Figures 3 and 5 are the fits of this theory to the data, where the shape of $H_{c2}(T)$ is determined by the ratio of the in-plane Fermi velocities $\eta=(v_2/v_1)^2$ for band 1 and 2, the mass anisotropy ratio $\varepsilon=m_{ab}/m_c$, and the Pauli pair-breaking parameter $\alpha = \pi k_B T_c m_{ab}/E_F m_e$ ($m_{ab}$ is the electron effective mass in the $ab$ plane, $m_e$ is the free electron mass, $E_F = m_{ab} v_1^2/2$, and bands 1 and 2 correspond to the hole and electron pockets of the FS, respectively). In a single-band superconductor, the condition $\alpha \gtrsim 1$ means that Pauli pair-breaking significantly changes the shape of $H_{c2}(T)$ resulting in the FFLO instability. As shown in Figure 3, the upward curvature of $H_{c2}^{\perp}(T)$ for strongly underdoped Ba-3 corresponds to the $s^{\pm}$ pairing for $\eta=0.01$ and $\alpha=0.02$. Here paramagnetic effects are negligible in band 1 where $\alpha_1 \sim \alpha << 1$, but essential in band 2 for which $\alpha_2 \sim \alpha/\eta=2$. For $H//ab$, the change in the curvature of $H_{c2}^{//}(T)$ in Figure 3 is attributed to strong mass anisotropy $\varepsilon \sim 0.02$, which increases $\alpha \rightarrow \alpha\varepsilon^{-1/2}$ and flattens $H_{c2}^{//}(T)$ at low T. For Ba-1 and Ba-2, $H_{c2}^{\perp}(T)$ can be described assuming a moderate band asymmetry for the parameters given in the caption. The fitting curves $H_{c2}(T)$ for the optimally doped Ba-1 at low temperatures extrapolate to ≈ 100T above the upper limit of our high field measurements. The fits suggest the appearance of the FFLO state below 10K for the optimally doped Ba-1 sample at H//ab.

The shapes of $H_{c2}(T)$ for FeSe$_{1-x}$Te$_x$ films shown in Figure 5 indicate strong Pauli pairbreaking consistent with the previous measurements.[8,10] The fits correspond to $\alpha_{\perp}$=1.2 for $H\perp ab$ and $\alpha_{//}=\alpha_{\perp}\varepsilon^{1/2}$ = 7 and $\varepsilon$≈0.03 for H//ab, for which the theory suggests the FFLO instability below 3-9K where the FFLO wave vector Q(T) appears spontaneously as shown in Figure 5. The principal importance of the FFLO instability for the interpretation of our experimental data is particularly evident in the case of H//ab. Here $H_{c2}(T)$ calculated without the FFLO instability (dashed line in Fig. 5) is inconsistent with the observed $H_{c2}(T)$ which exhibits a very steep raise near $T_c$, indicative of $\alpha >>1$ and a nearly linear temperature dependence at low T. By contrast taking the FFLO instability into account in Eqs. (1)-(3) enables us to explain the shapes of the $H_{c2}^{//}(T)$ curves observed on our FeSe$_{1-x}$Te$_x$ films, as shown in Fig. 5 (solid line). The conclusion about strong Pauli pairbreaking and the FFLO instability at H||ab is consistent with the parameters of FeSe$_{1-x}$Te$_x$ extracted from ARPES[27] which yields $E_F$≈20-30meV and $m_{ab}/m_e$ = 3-16 for different FS pockets, giving $\alpha$≈0.5-4 for $H\perp ab$. Next we use the orbitally limited slope $H_{c2}'$ near $T_c$ in the two-band model[12]

$$H_{c2}' = \frac{24\pi k_B^2 T_c \phi_0}{7\zeta(3)\hbar^2 (c_1 v_1^2 + c_2 v_2^2)} \qquad (4)$$

where $\zeta(3) = 1.202$, and $c_1$ and $c_2$ depend on inter and intraband coupling constants. For the s$^{\pm}$ model in which interband pairing dominates, $c_1 \to c_2 \to$ ½ and Eq. (3) yields $H_{c2}'[T/K]$ = $1.8T_c[K]m_{ab}/\{m_e(1+\eta)E_{F1}[meV]\}$. The observed slope $H_{c2}'$≈8 T/K corresponds to $m_{ab}/m_e$≈12 for $T_c$=18K, $E_F$=25meV and $\eta=1$ or $m_{ab}/m_e$≈6 if $\eta<<1$. Again these values of $m_{ab}/m_e$ are qualitatively consistent with the ARPES data.[27] Small $E_F$ and large $m_{ab}$, in FeSe$_{1-x}$Te$_x$ thus yield $\alpha>1$ even for *H//c,* so the FFLO state for *H//ab* is a realistic possibility which may be explicitly verified by magnetic torque, specific heat or NMR measurements. The very short GL coherence lengths $\xi_{ab} = (\phi_0/2\pi T_c H_{c2}')^{1/2} \approx 1.5$nm for $H_{c2}'$=8.1T/K ($H \perp ab$) and $\xi_c = \xi_{ab} H_{c2}^{//}/H_{c2}^{\perp} < 0.12$nm for $H_{c2}'$>100 T/K (*H//ab*) extracted from the data in Figure 5 show that our FeSe$_{1-x}$Te$_x$ films are indeed in the clean limit.

The multiband structure of FBS allows tuning $H_{c2}(T)$ by small shifts of the Fermi level, which changes the Fermi velocities and can open up small FS pockets which enhance the Pauli pair-breaking.[12] For instance, Ba$_{1-x}$K$_x$Fe$_2$As$_2$ has two or three hole FS pockets and two electron FS pockets.[17-20] In particular, as shown in ARPES experiments,[19,20] in the underdoped cases there is a hole band at Γ below the Fermi level but at the optimal doping this band emerges and cross the Fermi level. So doping changes the ratio of the Fermi velocities $\eta=(v_2/v_1)^2$ and the intraband parameters $\alpha$, which in turn changes the shape of $H_{c2}(T)$ in Figure 3. The $\alpha$ parameter for Ba-1 is significantly larger than for the two underdoped sample (0.5 against 0.02-0.05) so the effect of the Pauli pairbreaking on the shape of $H_{c2}$(T) curves in Ba-122 remains relatively moderate, unlike he more Pauli limiting case of FeSe$_{1-x}$Te$_x$.

An interesting situation occurs in FeSe$_{1-x}$Te$_x$ which has two hole FS pockets at the Γ point of the Brillouin zone, and a hole band just slightly below the Fermi level at *x* = 0.58 [28,27,29]. This band structure allows us to use the following model[12] of the extremely high $H_{c2}$ slopes in our strained FeSe$_{1-x}$Te$_x$ films. As *x* is changed from 0.58 ($T_c$=11.5K)[27] to 0.5 $T_c$ climbs up to 16K (18K under strain) and a new FS hole pocket emerges, while the electron FS pocket at M shrinks. This emerging hole pocket, driven either by doping or by strain, has a small Fermi energy $E_F$ and strongly enhanced effective mass $m_{ab}$[27,29,30] and thus a large Pauli pair-breaking parameter $\alpha \propto m_{ab}/E_F$, which reduces the role of orbital pair-breaking for the other hole FS pockets at the Γ point, thus increasing the slopes of $H_{c2}$ at $T_c$.[12] Consistent with this model, the large substrate-induced compressive strain of our films shows much higher $H_{c2}$ slopes than previous measurements on FeSe$_{1-x}$Te$_x$ single crystals,[8-10] and the huge slopes for H//*ab*. The emerging hole FS pocket can further reduce the small c-axis GL coherence length $\xi_c \approx 0.12$ nm which is already less than half the spacing $c = 0.55$ nm between the *ab* planes[2] for $H_{c2}'>100$ T/K for H//*ab* shown in Figure 5. The slopes $H_{c2}'>100$ T/K in our strained films are consistent with the collapse of $\xi_c$ triggered by the opening of the hole FS pocket which decouples the *ab* planes and results in a dimensional crossover,[31] $\xi_c(T)<2^{-1/2}c$ at $T = T^*$ very close to $T_c$. For $T < T^*$, $H_{c2}^{//}(T)$ is mostly limited by the Pauli pair-breaking for which $H_{c2}^{//}(T) \propto (T^* - T)^{1/2}$, where T* is very close to $T_c$. So the slope $H_{c2}'$ at $T_c$ is enhanced by the large factor $\approx [T_c/(T_c - T^*)]^{1/2}$ with respect to the orbital limit. Indeed $H_{c2}^{//}(T) \propto (T_c - T)^{1/2}$ describes very well our experimental data on the films with highest $H_{c2}$ slopes near $T_c$, as is displayed in Figure 5d.

## IV. DISCUSSION

To demonstrate how much $H_{c2}$ can be enhanced by doping and strain, we compare our optimally doped Ba-1 crystal and FeSe$_{1-x}$Te$_x$ films with other superconductors. Figure 6 shows that $H_{c2}(0)$ for FeSe$_{1-x}$Te$_x$ is almost twice higher than for Nb$_3$Sn despite their same $T_c$ = 18K. Yet the Pauli-limited $H_{c2}(0)$ of FeSe$_{1-x}$Te$_x$ is lower than $H_{c2}(0)$ of PbMo$_6$S$_8$[32] which has $T_c$ = 14 K and the orbitally-limited shape of $H_{c2}(T)$,[16] likely due to larger E$_F$ and smaller α. We would like to emphasise that our Ba$_{0.6}$K$_{0.4}$Fe$_2$As$_2$ single crystal has much higher $H_{c2}(T)$ than Ba$_{0.55}$K$_{0.45}$Fe$_2$As$_2$[4] and Ba$_{0.68}$K$_{0.32}$Fe$_2$As$_2$[21] whereas below 36K Ba$_{0.6}$K$_{0.4}$Fe$_2$As$_2$ surpasses NdFeAsO$_{1-x}$F$_x$[11] despite its higher $T_c$. The latter indicates that $H_{c2}$(T) in Fe-based superconductors do not simply scale with T$_c$ since the multiband effects and the interplay of orbital and Pauli pairbreaking can result in a more complex and interesting behaviour.

Our high field measurements have shown that the observed shape of $H_{c2}$(T) curves for H//*ab* in chalcogenide films is inconsistent with the conventional WHH-based multiband theory[12,16] if the FFLO instability is disregarded but it can be explained by taking into account the FFLO state at low temperatures. We hope that this result may motivate other experimental groups to verify it by independent NMR, magnetic torque or specific heat measurements. However, the two latter techniques may not unambiguously reveal the FFLO state if the weak features of the first order FFLO transition[15] are smeared by $T_c$ inhomogeneities characteristic of even the best chalcogenide single crystals.[26]

In conclusion, our high-field measurements show quite explicitly how the behaviour of $H_{c2}(T)$ in $Ba_{1-x}K_xAs_2Fe_2$ can be effectively tuned by doping resulting in $H_{c2}$ ~ 85T at ~$T_c$/2. Our strained $FeSe_{1-x}Te_x$ films exhibit an extreme Pauli-limited $H_{c2}(T)$, indicative of the Fulde-Ferrel-Larkin-Ovchinnikov state. The observed suppression of orbital pair-breaking causes significant enhancement of $H_{c2}$ of the optimally doped $Ba_{1-x}K_xAs_2Fe_2$ (at low temperature more than 25 T larger as compared to the previous results[4,21] on both slightly underdoped and overdoped $Ba_{1-x}K_xAs_2Fe_2$) which could make Ba-122 a competitive high-field magnet material.

**ACKNOWLEDGEMENTS**

A portion of this work was performed at the National High Magnetic Field Laboratory, which is supported by NSF Cooperative Agreement No. DMR-0654118, by the State of Florida, and by the DOE. The work in Beijing is supported partly by the Ministry of Science and Technology of China (973 Project: 2011CBA01000) and the Natural Science Foundation of China. We are grateful to M. Jaime, J. Betts and M. Putti for discussion and experimental help.

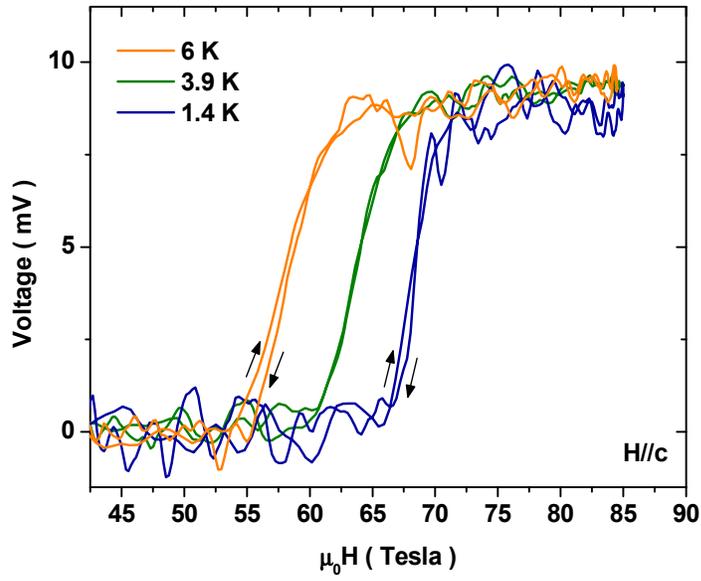

**Figure 1** (Color online) Superconducting transitions for Ba-2 measured in the 85T pulsed magnet for H//*c* showing both the up and down field sweeps.

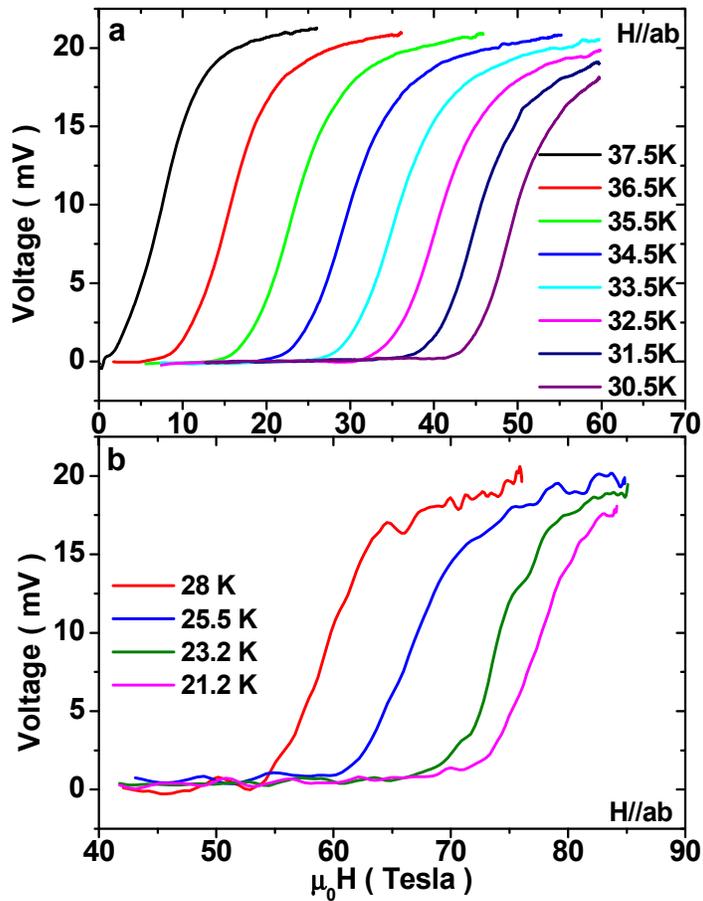

**Figure 2** (Color online) Superconducting transitions for Ba-1 in the 60T (a) and 85T (b) pulsed magnets.

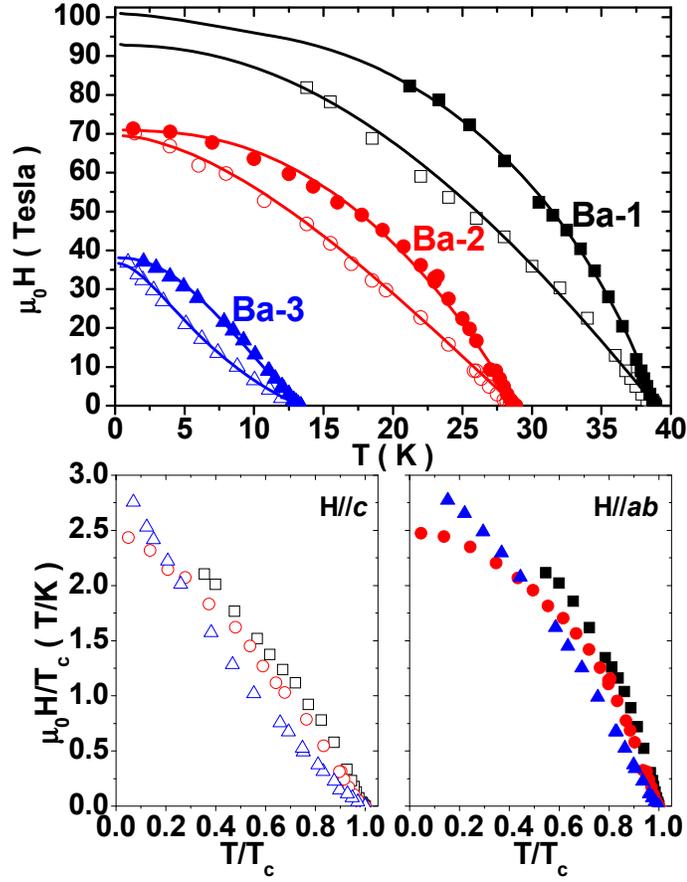

**Figure 3** (Color online) (a) $H_{c2}(T)$ curves for $Ba_{1-x}K_xAs_2Fe_2$ single crystals with different doping levels (H//*ab* and H//*c*, full and empty symbols respectively) and fits with the two-band theory (continuous lines). For Ba-1: s=1, $\eta$=0.9, $\alpha\varepsilon^{1/2}$=1.9 (H//*ab*) and s=1, $\eta$=0.9, $\alpha$=0.5 (H//*c*). For Ba-2: s=1, $\eta$=0.6, $\alpha\varepsilon^{1/2}$=0.8 (H//*ab*) and s=1, $\eta$=0.6, $\alpha$=0.05 (H//*c*). For Ba-3: s=1, $\eta$=0.07, $\alpha\varepsilon^{1/2}$=0.14 (H//*ab*) and s=1, $\eta$=0.01, $\alpha$=0.02 (H//*c*). (b) and (c) data from panel a normalized to $T_c$.

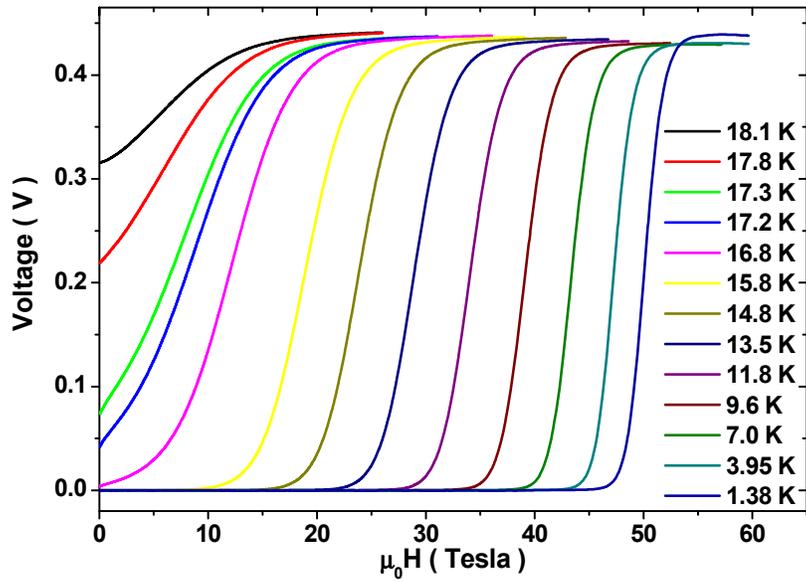

**Figure 4** (Color online) Superconducting transitions for FeSe$_{1-x}$Te$_x$ strained thin films in high field.

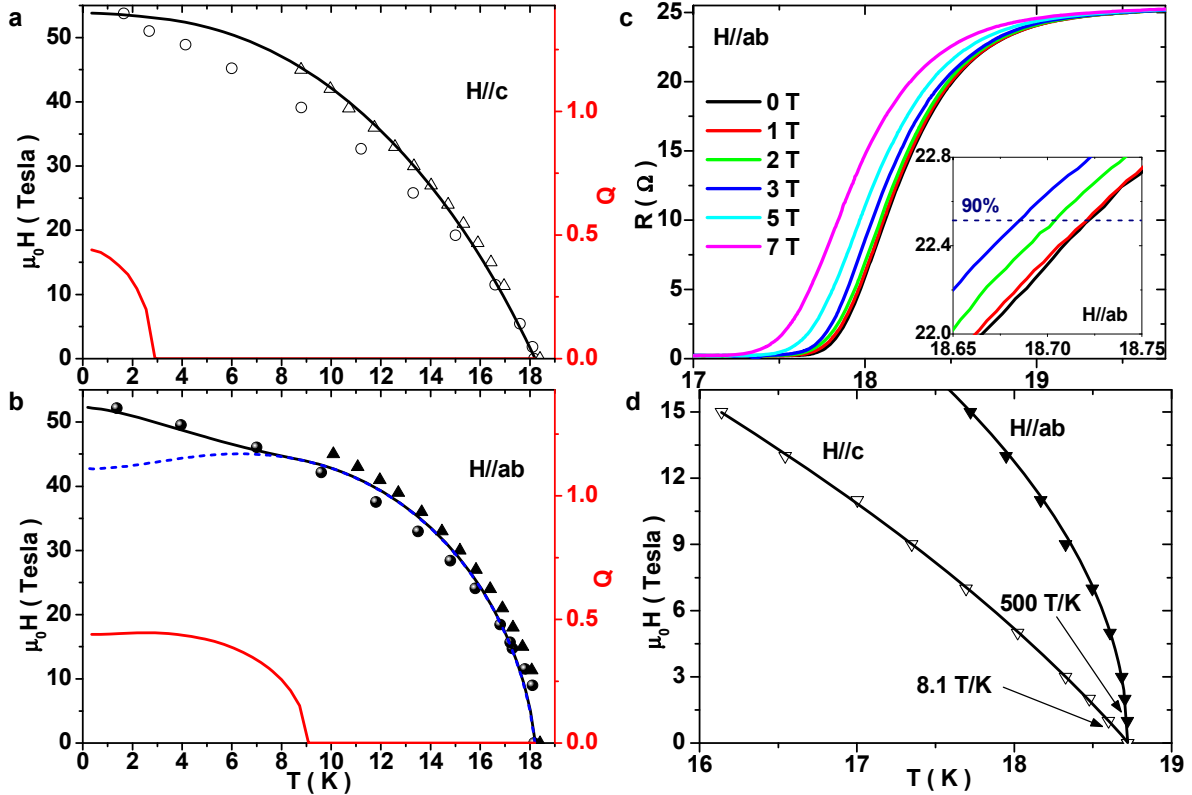

**Figure 5** (Color online) (a) and (b) $H_{c2}(T)$ curves for two FeSe$_{1-x}$Te$_x$ strained thin films with similar $T_c$ measured in the 45T DC magnet (triangles) and in the 60T pulsed magnet (circles) (full and empty symbols represent H//*ab* and H//*c*, respectively). The black solid curves correspond to $H_{c2}$ fits and the red curves to the Q vector of the FFLO state. The blue dashed line in (b) shows $H_{c2}(T)$ calculated without taking the FFLO instability into account. Fit parameters are: $\alpha$=1.2 and $\eta$=0.9 for H//*c* and $\alpha\varepsilon^{1/2}$=7 for H//*ab*. Here Q is measured in the units of $Q_0=4\pi k_B T_c/\hbar v_1 \varepsilon^{3/4}$. (c) Superconducting transitions at low field (H//*ab*) of a FeSe$_{1-x}$Te$_x$ film and magnification close to 90% of the normal state resistivity. (d) For the same sample, the detailed $H_{c2}$ data close to $T_c$ show the extremely high slope for H//*ab*. Here the continuous lines show the $H_{c2}(T) \propto (T_c - T)^{1/2}$ indicating a complete suppression of the orbital pairbreaking.

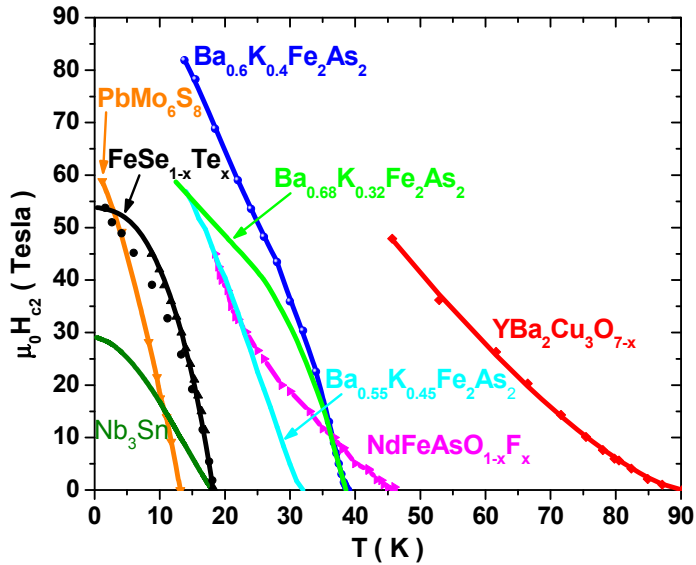

**Figure 6** Comparison of the upper critical fields of different materials: $H_{c2}$(T) data of Nb$_3$Sn, PbMo$_6$S$_8$,[32] FeSe$_{1-x}$Te$_x$ (this paper), Ba$_{0.55}$K$_{0.45}$Fe$_2$As$_2$,[4] Ba$_{0.68}$K$_{0.32}$Fe$_2$As$_2$,[21] Ba$_{0.6}$K$_{0.4}$Fe$_2$As$_2$ (this paper), NdFeAsO$_{1-x}$F$_x$ [11] and the irreversibility field of YBa$_2$Cu$_3$O$_{7-x}$.

[1] Y. Kamihara, T. Watanabe, M. Hirano and H. Hosono, J. Am. Chem. Soc. **130,** 3296 (2008).

[2] D.C. Johnston, Adv.Phys. **59**, 803 (2010).

[3] J. Paglione, and R.L. Greene, Nature Phys. **6,** 645 (2010).

[4] M. M. Altarawneh *et al*., Phys. Rev. B **78,** 220505(R) (2008).

[5] G. Fuchs *et al.*, New J. Phys. **11,** 075007 (2009).

[6] A. Yamamoto *et al*., Appl. Phys. Lett. **94,** 062511 (2009).

[7] H. Q. Yuan, *et al*., Nature **457,** 565 (2009).

[8] M. Fang *et al*., Phys. Rev. B **81,** 020509 (R) (2010).

[9] T. Kida *et al.*, J. Phys. Soc. Jpn. **78,** 113701 (2009).

[10] T. Klein *et al.,* Phys. Rev. B **82,** 184506 (2010).

[11] J. Jaroszynski *et al*., Phys. Rev. B **78,** 174523 (2008).

[12] A. Gurevich, Phys. Rev. B **82,** 184504 (2010).; Rep. Prog. Phys. **74**, 124501 (2011)

[13] P. Fulde and R. A. Ferrel, Phys. Rev. **135,** A550 (1964).

[14] A. I. Larkin and Y. N. Ovchinnikov, Zh. E ksp.Teor.Fiz. **47,** 1136 (1964) [Sov. Phys. JETP, **20,** 762 (1965)].

[15] Y. Matsuda and H. Shimahara, J. Phys. Soc. Jpn. **76,** 051005 (2007).

[16] N.R.Werthamer, E. Helfand, and P.C. Hohenberg, Phys. Rev. **147,** 295 (1966).

[17] Ding, H. *et al.*, Eur. Phys. Lett.**83**, 47001 (2008).

[18] M. Yi *et al.*, Phys. Rev. B **80,** 024515 (2009).

[19] H. Ding *et al.*, J. Phys.: Condens. Matter **23**, 135701 (2011).

[20] M. Neupane *et al.*, Phys. Rev. B **83**, 094522 (2011).

[21] V. A. Gasparov, F. Wolff-Fabris, D. L. Sun, C. T. Lin, and J. Wosnitza, JETP Letters **93**, 26 (2001).

[22] H. Q. Luo *et al.*, Supercond.Sci. Technol. **21,** 125014 (2008).

[23] G. Mu *et al.*, Phys. Rev. B **79,** 174501 (2009).

[24] E. Bellingeri *et al.*, Supercond.Sci. Technol. **22,** 105007 (2009).

[25] E. Bellingeri *et al.*, Appl. Phys. Lett. **96,** 102512 (2010).

[26] A. Serafin *et al.*, Phys. Rev. B 82, 104514, (2010).

[27] A. Tamai *et al.*, Phys. Rev. Lett. **104,** 097002 (2010).

[28] F. Chen *et al*., Phys. Rev. B **81,** 014526 (2010).

[29] K. Nakayama *et al*., Phys. Rev. Lett. **105**, 197001 (2010).

[30] H. Shishido *et al.*, Phys. Rev. Lett. **104**, 057008 (2010).

[31] R. A. Klemm, A. Luther, and M. R. Beasley, Phys. Rev. B **12**, 877 (1975).

[32] K. Okuda, M. Kitagawa, T. Sakakibara, M. Date, J. Phys. Soc. Jap. **48,** 2157 (1980).